\documentclass[doublecol]{epl2} 

\begin{document}

\title{Non-Hermitian Floquet invisibility}
\shorttitle{Non-Hermitian Floquet ... } 

\author{S. Longhi \inst{1,2}}
\shortauthor{S. Longhi}

\institute{                    
  \inst{1}  Dipartimento di Fisica, Politecnico di Milano, Piazza L. da Vinci 32, I-20133 Milano, Italy\\
  \inst{2}  Istituto di Fotonica e Nanotecnlogie del Consiglio Nazionale delle Ricerche, sezione di Milano, Piazza L. da Vinci 32, I-20133 Milano, Italy}


\abstract{Wave transport and scattering in open systems can be profoundly affected by non-Hermitian dynamics. In this work we consider wave scattering in a one-dimensional tight-binding lattice with a low-frequency harmonically-vibrating complex site potential. Floquet scattering is shown to be suppressed in a range of the spectral lattice band, which is limited by a singularity in the spectral transmission/reflection amplitudes for inelastic scattering channels. Invisibility over the entire spectral band is found when the singularity is pushed at the edge of the Brillouin zone, which requires a modulation frequency larger than the width of the tight-binding lattice band. Remarkably, invisibility is found to persist for multiple oscillating lattice impurities.}
\maketitle

\section{Introduction}
Over the last two decades, a great and increasing attention has been devoted to
study the properties of non-Hermitian systems \cite{r1,r2,r3,r4}. A wealth of important applications of non-Hermitian physics, such as those based on the existence of exceptional points \cite{r5,r6} or spectral singularities \cite{r7,r8,r9}, have been disclosed in the past few years in such diverse fields as optics, acoustics and optomechanics \cite{r10,r11,r12,r13,r14,r15,r16,r17,r18,r19,r20}.
Non-Hermitian Hamiltonians are generally used in both classical and quantum systems featuring energy dissipation and/or
gain. In such systems, wave transport, localization and scattering can be deeply modified as compared to Hermitian systems. Among the most intriguing effects found in non-Hermitian systems, we mention unidirectional or bidirectional invisibility \cite{r9,r12,r15,r21,r22,r23,r24}, non-Hermitian delocalization transition and robust transport in disordered systems \cite{r25,r26}, 
mobility transition from ballistic to diffusive transport \cite{r27}, hyperballistic transport \cite{r28}, chirality and unidirectional lasing \cite{r17,r18,r26,r28a}, breakdown of adiabatic theorem and topological energy transfer \cite{r19,r20,r27,r28b,r28c,r28d}, etc. The possibility to suppress wave scattering, thus realizing transparency effects in inhomogeneous media, is perhaps one among the most amazing phenomenon that occurs in non-Hermitian models. In previous studies \cite{r9,r12,r15,r21,r22,r23,r24,r28e,r28f,r28g}, invisibility has been observed in non-Hermitian systems described by time-independent Hamiltonians.\\
In this Letter we predict a novel kind of invisibility in wave scattering by a time-dependent oscillating non-Hermitian potential, an effect that can be referred to as {\em non-Hermitian Floquet invisibility}. The transmission of quantum or classical waves through a time-dependent
potential has been the subject of extensive studies
since more than three decades (see \cite{r29} and references therein). Floquet scattering and localization phenomena
 in modulated potentials involve fundamental aspects of quantum mechanics, such
as the problem of tunneling times \cite{r30,r31}, coherent control of tunneling \cite{r29}, classical and
quantum chaos \cite{r32,r33,r34}, and provide new ways to
manipulate wave transport in a wide variety of classical and quantum systems (see, for instance, \cite{r35,r36,r37} and references therein).
Important phenomena such as field-induced barrier transparency \cite{r38,r39}, quasi-bound states and Fano resonances \cite{r40,r41,r42,r43}, 
 high-frequency blockade states \cite{r40,r44,r45} and resonance catastrophe \cite{r46}, have been disclosed in such systems. Most of such previous studies, however, have been focused to Hermitian dynamics. Recently, it has been shown that Floquet scattering in a special class of oscillating non-Hermitian potential wells, synthesized by supersymmetric quantum mechanics, can behave in an unusual way,  leading to reflectionless and energy-conserving particle transmission \cite{r47}. However, such continuous potentials turn out to be rather exotic ones and their physical implementation is challenging.
 In the present Letter we consider Floquet scattering in a paradigmatic model of mesoscopic quantum or classical transport, namely Floquet scattering in a tight-binding lattice with a  low-frequency oscillating impurity site \cite{r42,r45,r46}. For non-Hermitian oscillation, i.e. involving oscillating gain and loss of the potential site, we show that invisibility can be observed in a spectral interval of the lattice energy band, which is bounded by the appearance of a singularity in the spectral transmission/reflection coefficients of inelastic scattering channels. Remarkably, invisibility persists in the presence of more than one oscillating non-Hermitian impurity in the lattice. 
 
\section{Floquet scattering in a tight-biniding lattice with an oscillating impurity site} 
Let us consider mesoscopic quantum or classical transport in a one-dimensional tight-binding lattice with an oscillating site impurity (Fig.1), which is described by the time-periodic Hamiltonian \cite{r45,r46}
\begin{equation}
\hat{H}= \sum_{n=-\infty}^{\infty} \kappa \left( |n \rangle \langle n+1| +  |n+1 \rangle \langle n | \right)+ V_0 f(t) |0 \rangle \langle 0|
\end{equation}
where $|n \rangle$ is the Wannier state localized at site $n$, $\kappa$ is the hopping rate between adjacent sites in the lattice, $V_0$ is the impurity potential at site $n=0$, and $f(t)$ describes harmonic oscillation in time of the impurity at frequency $\omega$. In the Hermitian case, $f(t)= \cos (\omega t)$. Here we extend the analysis to the case of non-Hermitian oscillation, namely we assume 
\begin{equation}
f(t)=\cos(\omega t)+i \Delta \sin (\omega t).
\end{equation}
 The Hermitian case, previously investigated in Refs.\cite{r42,r45,r46}, is obtained in the limit $\Delta=0$. For $\Delta \neq 0$, the impurity potential at site $n=0$ has a complex energy, and the real and imaginary parts of the  energy oscillate in phase quadrature at frequency $\omega$. A physical implementation of the Hamiltonian (1)  with non-Hermitian oscillation can be realized, for example, in photonic systems, such as in a chain of coupled microring resonators \cite{r42} where combined phase and amplitude modulation of the microring $n=0$ is impressed at frequency $\omega$. Another photonic system to implement the Hamiltonian (1) could be an array of evanescently-coupled optical waveguides in which the waveguide $n=0$ shows a modulation, along the longitudinal propagation direction $t$, of the real and imaginary parts of the refractive index \cite{r37}.
 
 \begin{figure}
\onefigure[width=8cm]{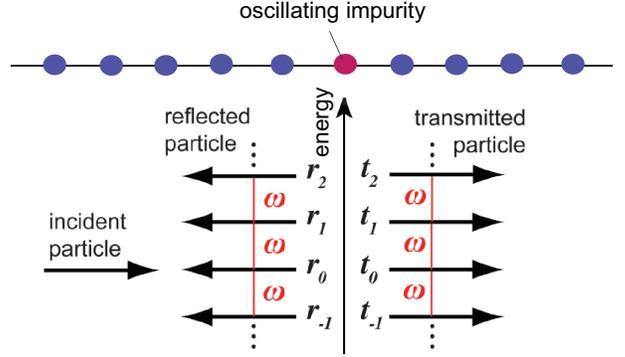}
\caption{(Color online) Schematic of Floquet scattering of a quantum particle propagating along a tight-binding lattice and scattered off by an oscillating impurity site.}
\end{figure}

 If the state vector $| \psi(t) \rangle$ of the system is expanded in series of the Wannier basis $|n \rangle$, $|\psi(t) \rangle=\sum_n c_n(t) |n \rangle$, the evolution of the amplitude probabilities $c_n(t)$ is governed by the following coupled equations  (with $\hbar=1$)
\begin{eqnarray}
i \frac{dc_n}{dt} & = & \kappa ( c_{n+1}+c_{n-1}) \nonumber \\
& + & \delta_{n,0}c_0  \left[ \Theta_1 \exp(i \omega t)+ \Theta_2 \exp(-i \omega t) \right]
\end{eqnarray}
where we have set $\Theta_1 \equiv (V_0/2) (1+ \Delta)$ and $\Theta_2 \equiv (V_0/2) (1- \Delta)$. In the static limit $V_0=0$, the lattice is homogeneous, scattering is prevented and Bloch states have the form $c_n(t)= \exp[-iqn-i E(q)t]$ with the dispersion relation $E(q)=2 \kappa \cos q$, where $-\pi \leq q < \pi$ is the Bloch wave number (quasi momentum). Forward-propagating waves in the lattice, corresponding to a positive group velocity $v_g= -(\partial E / \partial q)=2 \kappa \sin q$, have a Bloch wave number $q$ in the range $(0, \pi)$, whereas backward propagating waves correspond to $-\pi < q <0$. For $V_0 \neq 0$,  the site $n=0$ acts as a scattering center and Floquet theory should be applied to study wave scattering. The quasi-energy spectrum of $\hat{H}(t)$ is composed rather generally by scattering (delocalized) states and possible bound states localized near the impurity site $n=0$. The quasi energies of scattering states are real, whereas complex quasi energies could be found for bound states when $\Delta \neq 0$. To avoid the appearance of unstable secularly-growing modes, we will consider here parameter values where there are not bound states. As shown in the Appendix, $\hat{H}(t)$ does not sustain any bound state, and hence its quasi energy spectrum is entirely real, whenever $|\Delta| \leq 1$ and $\omega \leq 4 \kappa$ (low-frequency oscillation regime). To determine Floquet scattering states, note that for $V_0 \neq 0$ Bloch waves incident upon the impurity site $n=0$ will be
 partially reflected and partially transmitted  via elastic and inelastic processes, i.e. involving the absorption or emission of energy quanta from the oscillating field $f(t)$\cite{r40,r45} (Fig.1).  The scattering process is reciprocal, i.e. reflection and transmission are the same for left  and right incidence sides. Therefore, we can limit to consider the scattering process for a forward-propagating wave.  Transmission and reflection coefficients can be determined by application of Floquet theory \cite{r40}. Assuming that a forward-propagating light wave with Bloch wave number $q$ ($ 0<q< \pi$) and energy $E=E(q)=2 \kappa \cos q$ is incident onto the impurity from the left to the right side of the lattice, the exact scattered solution to Eqs.(3) has the form \cite{r42,r45}
\begin{equation}
c_n(t)= \left \{ 
\begin{array}{cc}
\sum_{\alpha= -\infty}^{\infty}  \left\{ \delta_{\alpha,0} \exp(-iq_{\alpha} n) + r_{\alpha}   \exp(iq_{\alpha} n) \right\} \\ \times  \exp(-i \Omega_{\alpha} t) \;\;\; (n \leq -1) \\
\sum_{\alpha= -\infty}^{\infty}  t_{\alpha}  \exp(-iq_{\alpha} n)   \exp(-i \Omega_{\alpha} t) \;\; (n \geq 0)
\end{array}
\right.
 \end{equation}
where $\Omega_{\alpha}= E(q) + \alpha \Omega$, $r_{\alpha}=r_{\alpha}(q)$ and $t_{\alpha}=t_{\alpha}(q)$ are the reflection and transmission amplitudes of the various Floquet (scattered) orders $ \alpha=0, \pm 1, \pm2, \pm 3, ...$ (Fig.1), and $q_{\alpha}$ are defined from the relation 
\begin{equation}
\cos q_{\alpha}=\cos q+ \alpha \frac{ \omega}{ 2 \kappa},
\end{equation}
 with $0 \leq q_{\alpha} \leq \pi$ if $q_{\alpha}$ is real (propagative states) and ${\rm Im}(q_{\alpha}) <0$ if $q_{\alpha}$ is complex (evanescent states). The transmittance $T(q)$ and reflectance $R(q)$ probabilities can be then calculated as \cite{r42}
\begin{equation}
T(q)= \sum_{\langle \alpha \rangle} \frac{v_{g \alpha}}{v_{g 0}} |t_{\alpha}|^2 \;, \;\;\; R(q)= \sum_{\langle \alpha \rangle} \frac{v_{g \alpha}}{v_{g 0}} |r_{\alpha}|^2
\end{equation}
where $v_{g \alpha}= 2 \kappa \sin q_{\alpha}$ is the group velocity at the Bloch wave number $q_{\alpha}$ and the symbol $ \langle ... \rangle$ means that the sum is extended over the indices $\alpha$ corresponding to propagative modes (i.e. $q_{\alpha}$ real).  The terms with $\alpha=0$ in the sums of Eq.(6), i.e. $T_0 \equiv |t_0|^2$ and $R_0 \equiv |r_0|^2$, correspond to elastic scattering, i.e. the energy of transmitted and reflected wave is not altered by the oscillating potential at site $n=0$. The other terms that contribute to the total transmittance and reflectance, namely $T_{\alpha}(q) \equiv (v_{g \alpha}/v_{g 0}) |t_{\alpha}|^2 $ and $R_{\alpha}(q) \equiv (v_{g \alpha}/v_{g 0}) |r_{\alpha}|^2 $, correspond to inelastic scattering channels, with transmitted and reflected waves with a higher (for $\alpha >0$) or lower (for $\alpha <0$)  energy amount $ \alpha \omega$ (Fig.1). In the Hermitian limit $\Delta=0$ one has $R+T=1$ owing to flux conservation, however in the non-Hermitian case  such a rule is generally violated. The transmission amplitudes $t_{\alpha}$ of various Floquet orders can be found from the solution of the difference equation
\begin{equation}
 \Theta_1  t_{\alpha+1} +  \Theta_2   t_{\alpha-1}-(2 i \kappa \sin q_{\alpha})  t_{\alpha} = -(2 i \kappa \sin q)   \delta_{\alpha,0}
\end{equation} 
The reflection amplitudes $r_{\alpha}$ of various Floquet orders can be then obtained from the relation 
\begin{equation}
r_{\alpha}(q)= t_{\alpha}(q)-\delta_{\alpha,0}.
\end{equation}
The solution to Eq.(7) can be formally written as
\begin{equation}
t_{\alpha}= \sum_{\beta} \left( \mathcal{M}^{-1} \right)_{\alpha, \beta} \sigma_{\beta}
\end{equation}
where $\sigma_{\beta}=-(2 i \kappa \sin q) \delta_{\beta,0}$ and $\mathcal{M}$ is the Floquet channel matrix defined by 
\begin{equation}
\mathcal{M}_{\alpha, \beta}  =   \Theta_1  \delta_{\alpha, \beta-1}+ \Theta_2  \delta_{\alpha, \beta+1}-(2 i \kappa \sin q_{\alpha}) \delta_{\alpha, \beta}.
\end{equation}

 \begin{figure}
\onefigure[width=8.8cm]{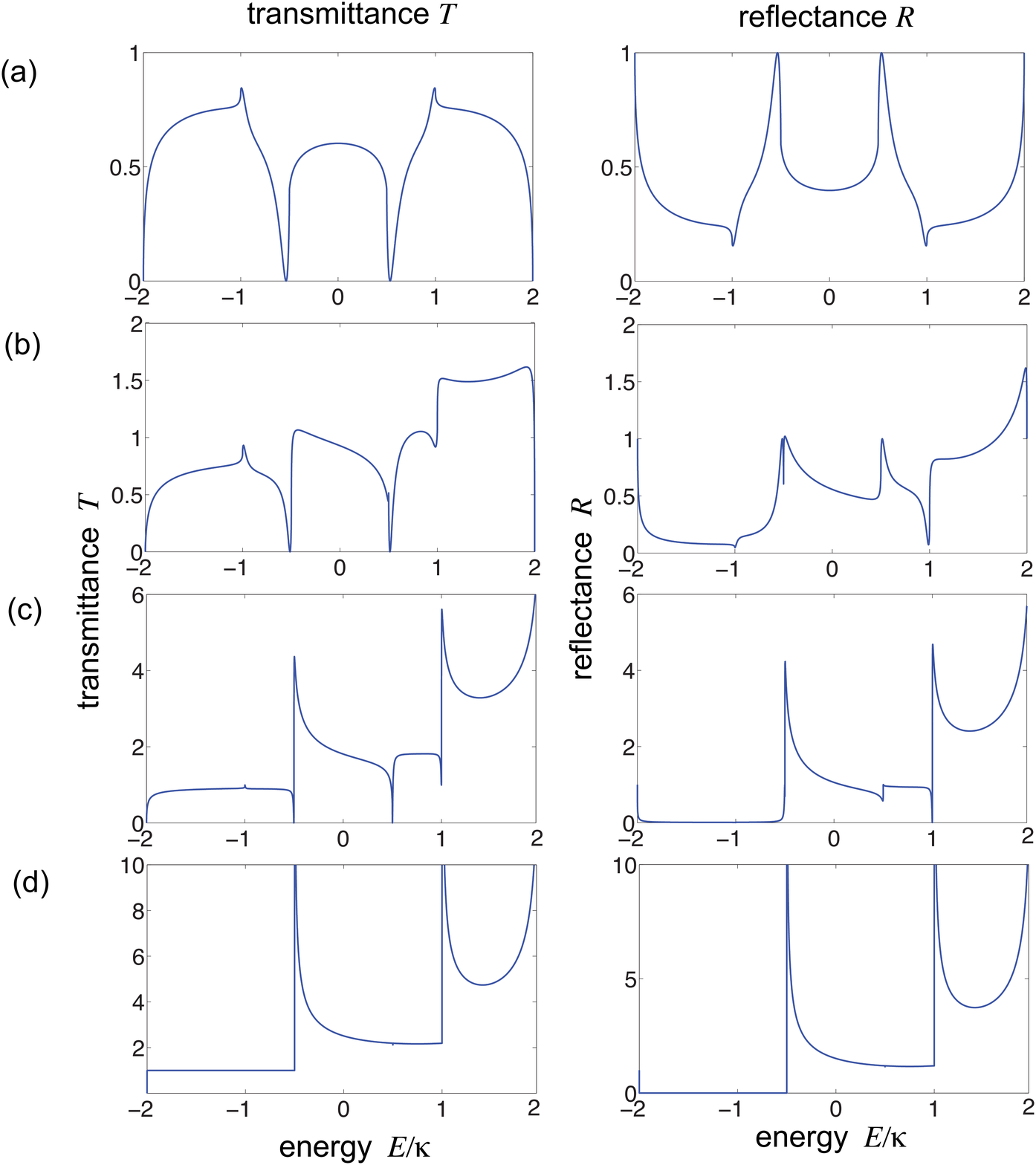}
\caption{(Color online) Numerically-computed total spectral transmittance $T$ (left panels) and reflectance $R$ (right panels) versus energy $E= 2 \kappa \cos q$ of incidence Bloch wave for parameter values $V_0 / \kappa=2$, $\omega/ \kappa=1.5$ and for increasing values of the non-Hermitian parameter $\Delta$: (a)  $\Delta=0$ (Hermitian limit), (b) $\Delta=0.6$, (c) $\Delta=0.9$ and (d) $\Delta=1$ (non-Hermitian invisibility).}
\end{figure}

\section{Non-Hermitian Floquet invisibility}
 In this study we focus our analysis to the most interesting case of low-frequency oscillation, corresponding to a modulation frequency $\omega$ smaller or equal the bandwidth $4 \kappa$ of the lattice band\footnote{ The high-frequency modulation regime $\omega \gg \kappa$ simply corresponds to a re-normalization of the hopping rate between the impurity site $n=0$ and its neighboring sites $n= \pm 1$. In this case, for $V_0$ of the order of $\sim \kappa$ (and thus $V_0 / \omega \ll 1$) transparency can be observed even in the Hermitian case $\Delta=0$. Such a transparency effect is however a trivial one and just arises because of the washing out of the impurity potential by the rapidly-oscillating field $f(t)$ (rotating-wave approximation).}. In the Hermitian limit $\Delta=0$, corresponding to $\Theta_1=\Theta_2=V_0/2$, the transmission spectrum shows Fano-like resonances for $\omega< 4 \kappa$, which have been investigated in previous works \cite{r42,r46}. An example of transmission and reflection spectra for $\Delta=0$ and for $\omega < 4 \kappa$, showing two resonance dips in the spectral transmission, is illustrated in Fig.2(a). As $\Delta$  slightly deviates from $\Delta=0$, a similar behavior is found, however as $\Delta$ in increased to approach $1$ from below a deeply different behavior is found, with the appearance of singularities in both spectral transmission and reflection curves at the energies $E=-2 \kappa+ \alpha \omega$ ($\alpha=1,2,...$)  inside the lattice band; see Figs.2(b-d). Remarkably, at $\Delta=1$ one has $T=1$, $R=0$ at energies $-2 \kappa<E<-2 \kappa+ \omega$, i.e. transparency is observed in such a spectral interval \footnote{A similar scenario is observed for $\Delta<0$ as $\Delta$ approaches -1 from above. In this case at $\Delta=-1$ invisibility is observed in the energy interval $(2\kappa-\omega,2 \kappa)$ rather than in the range $(-2 \kappa, -2 \kappa+\omega)$.}. As shown below, such a regime corresponds to {\em invisibility} in the spectral energy interval $(-2 \kappa, -2 \kappa+\omega)$, whereas the singularities observed in the spectral curves at energies $E=-2 \kappa+ \alpha \omega$ ($\alpha=1,2,...$) correspond to divergences of the inelastic scattering channels. Interestingly, as $\omega$ is increased toward $4 \kappa$ invisibility in the entire lattice energy band $(-2 \kappa, 2 \kappa)$ ca be achieved; see Fig.3.\\ The onset of invisibility in Floquet scattering can be rigorously proven by considering the form of the Floquet channel matrix $\mathcal{M}$ at $\Delta=1$. In this case $\Theta_1=V_0$, $\Theta_2=0$ and Eq.(7) can be solved in a closed form, yielding
 \begin{equation}
 t_{\alpha}(q)= \left\{
 \begin{array}{cc}
 0 & \alpha \geq 1 \\
 1 & \alpha=0 \\
 \frac{ V_0 }{ 2i \kappa \sin q_{\alpha} }t_{\alpha+1}=  \frac{ \pm i  V_0 }{ \sqrt{4 \kappa^2-(E+\alpha \omega)^2}}t_{\alpha+1}& \alpha \leq -1
 \end{array}
 \right.
 \end{equation}
 where we used the relation $2 \kappa \sin q_{\alpha}= \pm \sqrt{4 \kappa^2-(E+\alpha \omega)^2}$.
  \begin{figure}
\onefigure[width=8.8cm]{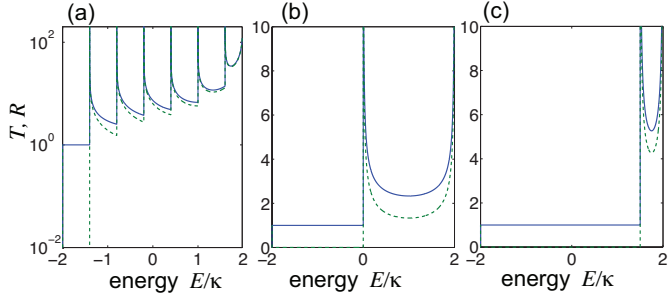}
\caption{(Color online) Numerically-computed total spectral transmittance $T$ (solid curves) and reflectance $R$ (dashed curves) versus energy $E$ of incidence Bloch wave for parameter values $V_0 / \kappa=2$, $\Delta=1$ and for (a) $\omega/ \kappa=0.6$, (b) $\omega/ \kappa=2$ and (c) $\omega/ \kappa=3.5$. Note that in (a) the vertical scale is logarithmic.}
\end{figure}
 The reflection coefficients $r_{\alpha}(q)$ at various Floquet channels are then obtained using Eq.(8). In particular, note that $r_0(q)=0$. This means that the impurity at site $n=0$ is fully invisible for the elastic scattering ($\alpha=0$) channel, i.e. $t_0(E)=1$ and $r_0(E)=0$ for any energy $E= 2 \kappa \cos q$ of the incident wave. However, for $\omega < 4 \kappa$ some of the inelastic channels $ \alpha=-1,-2,...$ may contribute to the total transmission and reflection [see Eq.(6)]. For an incidence wave with energy $-2 \kappa <E< -2 \kappa+\omega$, all inelastic scattering channels $\alpha=-1,-2,...$ refer to evanescent states and therefore do not contribute to the scattering process. This means that the wave is fully transmitted across the impurity site as if the impurity would not be present at all: this effect can be referred to as {\em non-Hermitian Floquet invisibility}. For an incidence wave with energy $E$ in the range $ -2 \kappa+\omega <E< -2 \kappa+ 2 \omega$, besides $\alpha=0$ also the inelastic channel $\alpha=-1$ contributes to wave scattering. Note that from Eq.(11) the transmission amplitude $t_{-1}$ shows a singularity (divergence) at the energy $E=-2 \kappa+\omega$. Likewise, for an incident particle with energy $E$ in the range $ -2 \kappa+2\omega <E< -2 \kappa+ 3 \omega$, besides $\alpha=0,-1$  also the inelastic channel $\alpha=-2$ contributes to wave scattering. Its amplitude
  $t_{-2}$ shows a singularity at the incidence energy $E=-2 \kappa+2 \omega$. \footnote{It should be noted that the divergences of the spectral amplitudes $t_{\alpha}$, $r_{\alpha}$ for $\alpha \leq -1$ are regularized, i.e. they actually do not lead to infinite transmission or reflection, when considering the propagation of a more physical normalizable wave packet, obtained by a superposition (integral) of scattering Bloch waves around some carrier energy $E=E_0$. For example, the singularity of  $t_{-1}(E)$ at $E=E_1=-2 \kappa+ \omega$ is of the type $t_{-1}(E) \sim (E-E_1)^{-1/2}$, which yields a finite integral when integrated around $E=E_1^+$. Scattering of wave packets with finite norm are shown in Figs.4 and 5.} The reasoning can be iterated to higher-order inelastic Floquet channels with any order $\alpha<0$ such that $-2\kappa-\alpha \omega$ remains smaller than $ 4 \kappa$. For example, in Fig.3(a) ($\omega/ \kappa=0.6$) there are six singularities in the spectral transmission and reflection curves at the energies $E=-2 \kappa- \alpha \omega$ with $\alpha=-1,-2,-3,-4,-5,-6$. 
  \par
  Non-Hermitian invisibility is expected to occur in the presence of more than one oscillating impurity site. In fact, let us assume $\hat{H}(t)=\hat{H}_0+f(t) \hat{H}_1$, where $\hat{H}_0=\kappa \sum_n (|n \rangle \langle n+1|+ | n+1 \rangle \langle n|)$ in the Hamiltonian of the homogeneous one-dimensional lattice, $\hat{H}_1=\sum_n V_n |n \rangle \langle n |$, $V_n$ is the potential energy at impurity site $|n \rangle$ (with $V_n=0$ for $|n|$ large enough), and $f(t)$ is the non-Hermitian oscillation defined by Eq.(2).  A Floquet scattering state of $\hat{H}(t)$, corresponding to an incident Bloch wave with energy $E= 2 \kappa \cos q$, can be written as $| \psi(t) \rangle=\sum_{\alpha=-\infty}^{\infty}| \Phi_{\alpha} \rangle \exp(-i \Omega_ \alpha t)$, where $\Omega_{\alpha}=E+ \alpha \omega= 2 \kappa \cos q_{\alpha}$. The asymptotic behavior of the  
Floquet components $| \Phi_{\alpha} \rangle$ as $n \rightarrow \pm \infty$ is similar to the one defined in Eq.(4), i.e.   $| \Phi_{\alpha} \rangle \sim \sum_n [\exp(-i q_{\alpha}n) \delta_{\alpha,0} + r_{\alpha} \exp(i q_{\alpha}n) ] |n \rangle$ for $n \rightarrow -\infty$ and $| \Phi_{\alpha} \rangle \sim \sum_n t_{\alpha} \exp(-i q_{\alpha}n)  |n \rangle$ for $n \rightarrow \infty$, where $r_{\alpha}$ and $t_{\alpha}$ are the spectral reflection and transmission amplitudes of the various Floquet channels. From the Schr\"{o}dinger equation, it follows that the   
  Floquet components $| \Phi_{\alpha} \rangle$ satisfy the recurrence relation
 \begin{equation}
 \Omega_{\alpha} | \Phi_{\alpha} \rangle= \hat{H}_0 | \Phi_{\alpha} \rangle + \Theta_1 \hat{H}_1 | \Phi_{\alpha+1} \rangle +\Theta_2 \hat{H}_1 | \Phi_{\alpha-1} \rangle 
 \end{equation} 
  where we have set $\Theta_1=(1+\Delta)/2$ and $\Theta_2=(1-\Delta)/2$. Let us assume $\Delta=1$, corresponding to $\Theta_1=1$ and $\Theta_2=0$. In this case the solution to Eq.(12) is given by $ | \Phi_{\alpha} \rangle=0$ for $ \alpha \geq 1$, $\hat{H}_0 | \Phi_0 \rangle=E | \Phi_0 \rangle$, and $| \Phi_{\alpha} \rangle=(\Omega_{\alpha}-\hat{H}_0 )^{-1} \hat{H}_1 | \Phi_{\alpha+1} \rangle$ for $\alpha \leq -1$. From the equation $\hat{H}_0 | \Phi_0 \rangle=E | \Phi_0 \rangle$ it follows that $t_{\alpha}(q)=1$ and $r_{0}(q)=0$, i.e. for the elastic scattering channel ($\alpha=0$) the impurity sites are fully invisible. For an energy $E$ of the incident wave in the range $ -2 \kappa < E < - 2 \kappa + \omega$, the only channel that contributes to scattering is the elastic channel $\alpha=0$, i.e. in Eq.(6) the sum is extended to the $\alpha=0$ index solely. Therefore in such an energy interval non-Hermitian Floquet invisibility is observed. Such a general result indicates that an arbitrary distribution $V_n$ of impurities in the lattice can be made invisible when they are collectively oscillating with complex amplitude  $f(t)= \exp( \pm i \omega t)$. 
\begin{figure}
\onefigure[width=8.8cm]{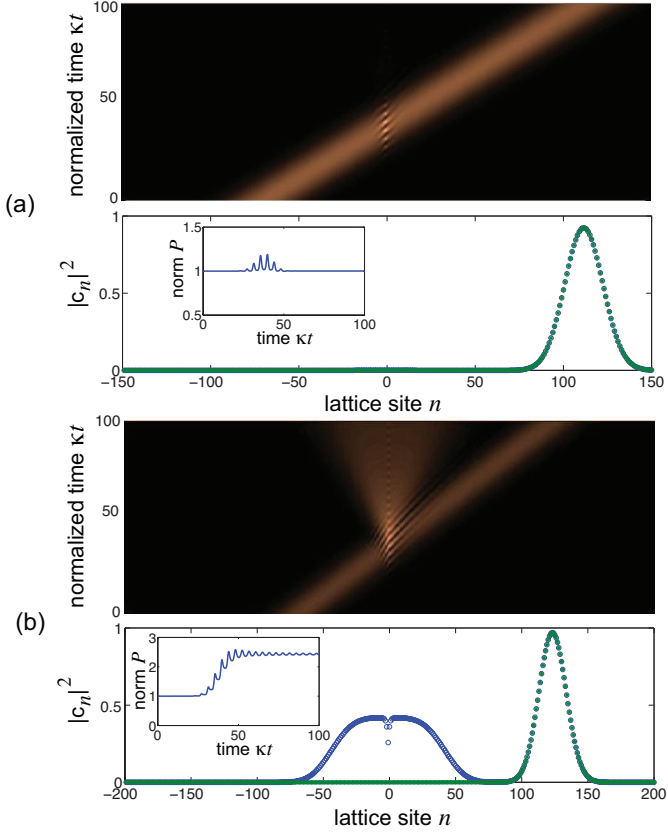}
\caption{(Color online) Propagation of a Gaussian wave packet of width $w=20$ across an oscillating impurity site at $n=0$ for parameters values $V_0/ \kappa=1$, $\omega / \kappa=1.5$ and $\Delta=1$. The carrier wave number $q_0$ of the wave packet is $q_0=2$ in (a), corresponding to an energy $E_0 / \kappa \simeq -0.83$ inside the invisibility range, and $q_0=1.8235$ in (b), corresponding to an energy $E_0 / \kappa=-0.5$ at the boundary of the transparency range. The upper pseudo color maps show the evolution of the amplitude probabilities $|c_n(t)|$ in the $(n,t)$ plane, whereas the lower plots depict the probability distribution $|c_n|^2$ at time $t=100 / \kappa$, i.e. after the scattering process (open circles). The asterisks curves in the plots show, for comparison, the behavior of the transmitted wave packet in the absence of the impurity. The insets depict the temporal evolution of the norm $P(t)=\sum_n |c_n(t)|^2/ \sum_n |c_n(0)|^2$.}
\end{figure}
  
  \section{Wave packet propagation}
  We checked the onset of non-Hermitian invisibility by direct numerical simulations of wave packet propagation in the tight-binding lattice with oscillating impurity sites. The time-dependent Schr\"{o}dinger equation was numerically integrated assuming as an initial condition a Gaussian wave packet with carrier Bloch wave number $q_0$ and width $w$, incident from the left side of the oscillating impurity sites. Typical numerical results are shown in Fig.4 for an impurity site at $n=0$ and for parameter values $V_0 / \kappa=1$, $\omega / \kappa=1.5$ and $\Delta=1$. In Fig.4(a) the energy $E_0=2 \kappa \cos q_0$ of the incident wave packet falls inside the spectral region $(-2 \kappa, -2 \kappa+ \omega)$ of invisibility, and thus the Gaussian wave packet is not scattered off by the oscillating impurity and propagates like in a homogeneous lattice [see the lower plot in Fig.4(a)]. In Fig.4(b) the energy $E_0$ of the the incident wave packet is taken at the singularity $E_0=E_1= -2 \kappa+ \omega$ of the $\alpha=-1$ inelastic scattering channel, i.e. at the boundary of the invisibility region. In this case, besides the elastic scattering process with unit spectral transmission and zero reflection [$t_0(E)=1$ and $r_0(E)=0$], leading to the main transmitted wave packet of Fig.4(b), scattering states corresponding to the $\alpha=-1$ channel are clearly observed after the interaction with the vibrating impurity site (the broadened wave packet spreading around $n=0$).\par
 Non-Hermitian transparency is observed also in the presence of more than one impurity site. As an example, Fig.5 shows the scattering of a Gaussian wave packet across a sequence of five lattice impurities of the same amplitude $V_0$, i.e. $V_n=V_0=$ for $|n|  \leq 2$ and $V_n=0$ for $|n| \geq 3$. Parameter values used in the simulations are $V_0/ \kappa=1$, $\omega / \kappa=1.5$ and $\Delta=1$. For an energy $E_0$ of the wave packet internal to the transparency interval [Fig.5(a)], scattering is not observed and the five impurities appear to be fully invisible [Fig 5(a)],while scattering is observed when the energy $E_0$ is tuned at the boundary or outside the invisibility region [Fig.5(b)]. 
     \begin{figure}
\onefigure[width=8.8cm]{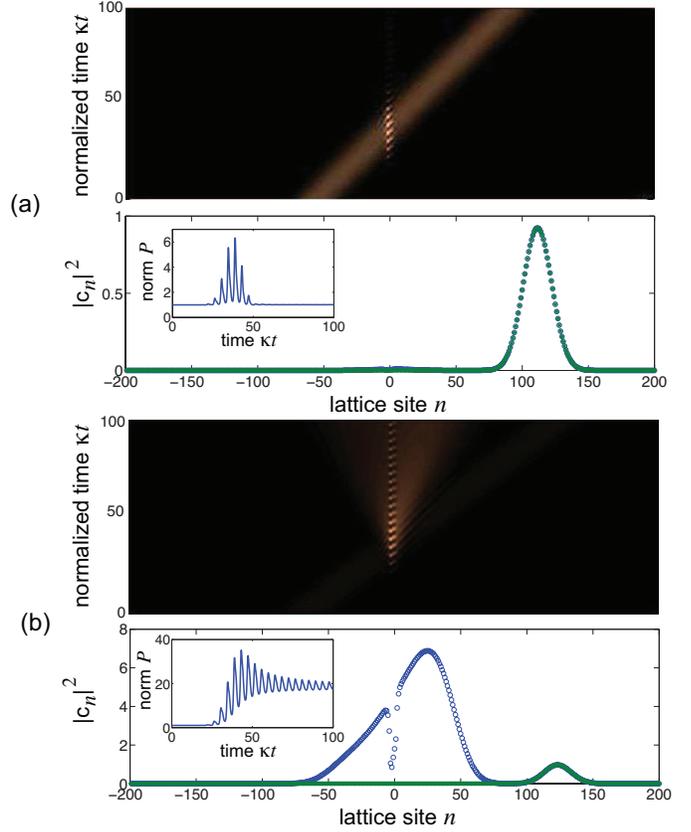}
\caption{(Color online) Same as Fig.4, but for a lattice with five collectively oscillating impurities of the same amplitude at sites $n=0, \pm1, \pm2$.}
\end{figure}



\section{Conclusions} 
The possibility to realize transparent or even invisible inhomogeneities in non-Hermitian classical and quantum wave transport is a fascinating property that has received a great attention in the past few years \cite{r10,r12,r15,r21,r22,r23,r24,r28f,r28g}. In most of such previous studies transparency was found for static (i.e. time independent) scattering potentials with specially-tailored spatial shape. For example, complex scattering potentials whose real and imaginary parts are related each other by spatial Kramers-Kronig relations are unidirectionally or bidirectionally transparent potentials \cite{r23,r24}.
In this work we have predicted a novel type of non-Hermitian bidirectional invisibility, that manifests itself as the absence of Floquet scattering of wave packets crossing oscillating complex impurity sites in a tight-binding lattice. The invisibility window can be tuned by varying the oscillation frequency of the impurities, with full  invisibility over the entire energy spectrum for a modulation frequency equal or larger than the lattice bandwidth. Remarkably, invisibility is found for arbitrary spatial distributions of the impurities when they are collectively oscillating with complex amplitude $f(t)=\exp( \pm i \omega t)$. Our results are expected to stimulate further theoretical and experimental studies in the rapidly emerging field of non-Hermitian classical and quantum transport. For example, non-Hermitian Floquet invisibility is expected to arise in continuous systems as well and could provide a viable route to make any arbitrary scattering potential invisible by oscillating it in time. 

\renewcommand{\theequation}{S-\arabic{equation}}
\setcounter{equation}{0}
\section{Appendix}

Bound states of $\hat{H}(t)$, localized near the impurity site $n=0$ and with quasi energy $E$, are solutions to Eq.(3) of the form
\begin{equation}
c_n(t)=\sum_{\alpha= -\infty}^{\infty} B_{\alpha} \exp(-iq_{\alpha} |n| -i \Omega_{\alpha} t)
 \end{equation}
where $\Omega_{\alpha}=E+ \alpha \omega$ and the complex numbers $q_{\alpha}$ are defined via the relation $ 2 \kappa \cos q_{\alpha}=\Omega_{\alpha}$. Localization requires ${\rm Im}(q_{\alpha}) <0$ for any index $\alpha$, i.e. for any Floquet scattering channel (including $\alpha=0$). Substitution of the Ansatz (S-1) into Eq.(3) yields the following recurrence relation for the amplitudes $B_{\alpha}$
\begin{equation}
\Theta_1 B_{\alpha+1}+ \Theta_2 B_{\alpha-1}-(2 i \kappa \sin q_{\alpha}) B_{\alpha}=0
\end{equation}
i.e. $\sum_{\beta} \mathcal{L}_{\alpha, \beta} B_{\beta}=0$, where the matrix $\mathcal{L}$ is defined by
\begin{equation}
\mathcal{L}_{\alpha,\beta}= \mp i \sqrt{4 \kappa^2-(E+ \alpha \omega)^2} \delta_{\alpha,\beta}+\Theta_{1} \delta_{\alpha, \beta-1}+ \Theta_2 \delta_{\alpha, \beta+1}
\end{equation}
and where we used the relation $ 2 \kappa \sin q_{\alpha}= \pm \sqrt{4 \kappa^2-(E+ \alpha \omega)^2}$. Since we require $B_{\alpha} \rightarrow 0$ as $| \alpha| \rightarrow \infty$, the index $\alpha$ can be truncated at some (possibly large) order, so that the solvability condition of Eq.(S-2) requires the vanishing of the determinant of the  matrix $\mathcal{L}^{(N)}$, which is obtained from $\mathcal{L}$ after truncation and $N$ is the number of Floquet amplitudes $B_\alpha$ included in the analysis. Since $\mathcal{L}^{(N)}$ is a tridiagonal matrix, its determinant $I_N$ can be calculated by a simple iterative map, and turns out to be a function of the quasi energy $E$, i.e. $I_N=I_N(E)$. A root $E=E_0$ of $I_N(E)$ defines a Floquet bound state provided that ${\rm Im}(q_{\alpha})<0$ for all $\alpha$  with non-vanishing amplitude $B_{\alpha}$. Let us first consider the Hemitian limit $\Delta=0$, so that $\Theta_1=\Theta_2=V_0/2$. Since the quasi energy spectrum of a time-periodic Hermitian Hamiltonian is entirely real, any root $E_0$ of $I_N(E)$ that corresponds to a bound state should be real. Since $q_{\alpha}$ is defined by the relation $2 \kappa \cos q_{\alpha}=E_0+ \alpha \omega$, for $\omega \leq  4 \kappa$  (the low-frequency oscillation regime considered in this work) there will be an index $\alpha=\alpha_0$ such that $-2 \kappa \leq E_{0}+ \alpha_0 \omega \leq 2 \kappa$. Since $q_{\alpha_0}$ is real, the root $E=E_0$ can not be accepted  because the localization condition fails. This means that in the Hermitian limit $\Delta=0$ and for $\omega \leq 4 \kappa$ there are not bound states, regardless of the value of $V_0$. Let us now consider the non-Hermitian regime with $|\Delta|<1$. In this case $\Theta_{1}$ and $\Theta_2$ are positive but different each other. Since $I_N$ depends on $\Theta_1$ and $\Theta_2$ solely via the product $\Theta_1 \Theta_2=V_0^2(1-\Delta^2)/4$, the roots of the equation $I_N(E)=0$ for $|\Delta|<1$ can be mapped into the ones of an Hermitian problem with renormalized amplitude $V_0'=V_0 \sqrt{1-\Delta^2}$, i.e. with the same value of $\Theta_1 \Theta_2$. Hence for $|\Delta|<1$ and in the low-frequency oscillation regime $\omega \leq 4 \kappa$ there are not Floquet bound states in the non-Hermitian case as well. Finally, let us consider the case $\Delta= \pm 1$, which corresponds to the invisibility regime considered in the main text. Let us assume, for example, $\Delta=1$, however a similar analysis holds for $\Delta=-1$. For $\Delta=1$ one has $\Theta_1=V_0$, $\Theta_2=0$, the matrix $\mathcal{L}$ has a bock diagonal form and its eigenvalues $\lambda_{\beta}$ are the elements of the main diagonal, i.e.  $\lambda_{\beta}=\mp i \sqrt{4 \kappa^2-(E+\beta \omega)^2}$. Since the determinant of the matrix $\mathcal{L}$ is the product of its eigenvalues, the roots $E$ of the determinantal equation ${\rm det} \mathcal{L}(E)=0$ are given by $E=-\beta \omega \pm 2 \kappa$ ($\beta=0, \pm 1, \pm 2, ...$).  For one of such values of $E$, say $E=E_{\beta}$ corresponding to some index $\beta$, the solution $B_{\alpha}^{(\beta)}$ to Eq.(S-2) reads $B_{\alpha}^{(\beta)}=0$ for $\alpha > \beta$, $B_{\beta}^{(\beta)}=1$, and $ B_{\alpha}^{(\beta)}= \mp i \Theta_1 B_{\alpha+1}^{(\beta)}/ \sqrt{4 \kappa^2-(-\beta \omega \pm 2 \kappa +\alpha \omega)^2}$  for $\alpha < \beta$. Since $q_{\beta}=0, \pm \pi$ and $B_{\beta}^{(\beta)}  \neq 0$, the corresponding solution (S-1) is however not bounded around $n=0$, and hence $E=E_{\beta}$ does not belong to the quasi-energy spectrum of $\hat{H}(t)$. Therefore also for $\Delta= \pm1$ $\hat{H}(t)$ does not sustain Floquet bound states.

\end{document}